# Local Proton Disorder Induced Intermolecular H-H Coupling in Ionization of Dense Ammonia

Yu Tao, Li Lei*, Jingyi Liu and Binbin Wu

*Institute of Atomic and Molecular Physics, Sichuan University, 610065 Chengdu, China*

*Electronic mail: lei@scu.edu.cn

**Abstract:** Under cold compression, hydrogen bonding (N–H···N) was considered to dominate intermolecular interaction during the ionization of ammonia. Here, we provide experimental and theoretical evidence of intermolecular H-H (N–H···H–N) coupling in dense ammonia. *Ab initio* molecular dynamics simulations (AIMD) reveal an increasing degree of proton disorder in ammonia with increasing pressure, which promotes intermolecular H–H coupling. At pressure below the ionization phase transition pressure (∼135 GPa), intermolecular H-H coupling gives rise to a unique dissociation pathway in ammonia. Sporadic molecular hydrogen was observed without laser heating, accompanied by the absence of molecular nitrogen. At pressure above ∼ 135 GPa, intermolecular H-H coupling serves as an intermediate state in the ionization of ammonia. Two fingerprint Raman modes ($L_1$ and $v_b$) previously assigned to the ionic phase (β) disappear upon further compression or heating. Together with proton transfer based on hydrogen bond, a dual-path mechanism exists in the ionization of ammonia. Our results demonstrate a case of hydride's phase transition pathway that occurs independently of hydrogen bonding under high pressure.

Molecular crystals are a unique class of solid-state matter, exhibiting a series of complex phase transition behaviors under high pressure. Namely, molecular hydrogen is predicted to transform into a metallic or atomic state [1–4], while molecular nitrogen can form polymeric nitrogen under high pressure and high temperature [5–7]. The nature of intermolecular interactions plays a crucial role in governing the high-pressure phase transition behavior of molecular crystals. Understanding the phase transition mechanism of molecular crystals under high pressure is a fundamental issue in condensed matter physics. Notably, HB-type hydrides ($NH_3$, $H_2O$, $H_2S$) display a unique dissociation mechanism under high pressure [8–10]. Their intermolecular interactions are dominated by hydrogen bonding. Hydrides can form a distinct ionic state under high pressure due to the presence of hydrogen bonds. Ammonia and water are believed to be major components of the interiors of Uranus and Neptune. Their ionic state could be associated with the existence of exotic magnetic fields in Uranus and Neptune, making them highly significant in planetary science [11–15].

Experimental evidence has been found for the transition of ammonia from solid molecular phases to an ionic phase [16,17]. Within 200 GPa over 0-3500 K temperature range, phases I-V are molecular phases [18–23], while phases α and γ are superionic phases [24,25], and β is the ionic phase. For molecular phases, phases IV and V are isostructural, differing only in the slight alteration of hydrogen bond angles [26]. The existence of ionic ammonia was confirmed by both Raman spectroscopy and synchrotron X-ray diffraction (S-XRD). However, the discrepancies between experimental and theoretical results indicate that our understanding of the phase transition mechanism of ammonia remains incomplete. Firstly, the key characteristics of the phase β measured by Raman spectroscopy are absent from the theoretically predicted stable structure with the *Pma*2 or a mixed phase with the *Pca*$2_1$. Additionally, ammonia exhibits a stable phase V structure over a broad pressure range from 18 to 150 GPa. Its vibrational mode exhibited a slight change around 70 GPa. This change was previously identified as a transition to phase VI [27]. The existence of phase VI has not been subsequently confirmed by S-XRD [28]. S-XRD cannot effectively capture the behavior of hydrogen in ammonia due to the very weak diffraction peak from hydrogen [29,30]. The behavior of hydrogen atoms may serve as the key factor responsible for the controversy.

This letter aims at revealing the role of hydrogen atoms in the phase V to β transition to resolve the existing experimental and theoretical discrepancies. We performed an in-depth analysis of Raman spectra up to 200 GPa. A total of seven runs were conducted. We found a unique dissociation pathway in dense ammonia. Theoretical calculations were performed to determine the intermolecular coupling of ammonia molecules. The unique dissociation pathway is attributed to the intermolecular H–H coupling induced by local proton disorder. It is also supported by a comparative analysis of softening vibrational modes in dense hydrogen, nitrogen and ammonia. Their relationship can be well described by an equation of the form $\sqrt{Z_\mathrm{H}/Z_N} = k_{\mathrm{N-N}}/k_{\mathrm{H-H}}$. Our results suggest that the ionization mechanism of ammonia based static structure, which relies on proton transfer along hydrogen bonds, is incomplete.

Fig.1 illustrates two possible ionization pathways of ammonia. The blue pathway represents the mechanism inferred from the static structure model, based on hydrogen bonding. There is a possibility that the orientations of ammonia molecules are not fixed due to the quantum effects of hydrogen or external perturbations. The green path schematically illustrates the H-H coupling arising from variations in molecular orientation. It could serve as an intermediate state toward ionization. With increasing pressure, the proton eventually transfers to another ammonia molecule, resulting in ionic ammonia. Briefly, proton transfer via hydrogen bonds is a one-step process, whereas H-H coupling proceeds in two steps. Our results support the existence of intermolecular H-H coupling in dense ammonia, besides hydrogen bonding. A dual-path mechanism exists in ionization of ammonia.

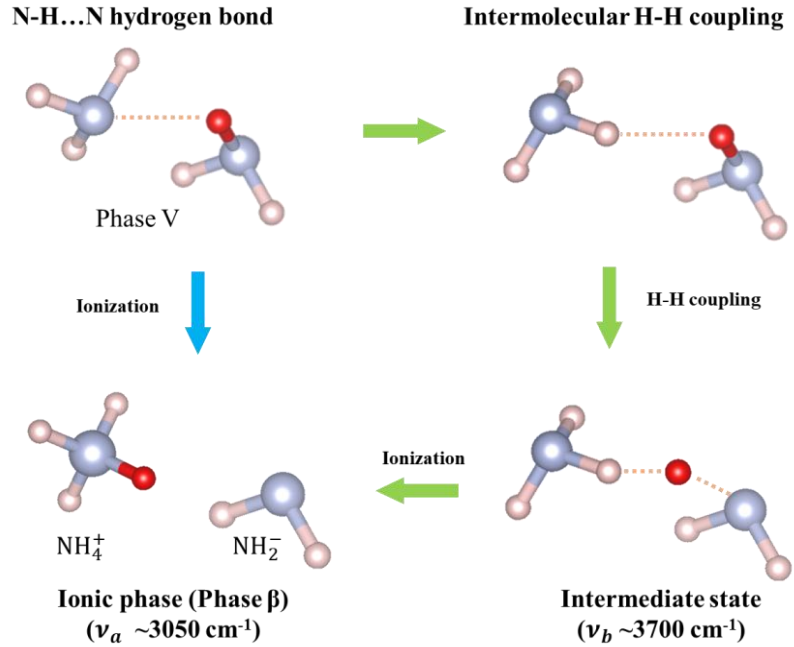

Fig. 1. Two possible ionization pathways of ammonia are illustrated. The blue arrow represents proton transfer via hydrogen bonds. The green arrow represents proton transfer via intermolecular H-H coupling. The ionization phase transition pressure is ~ 135 GPa, associated with $v_a$. The $v_b$ mode associated with an intermediate state disappears upon further compression or heating [Fig. 2 (d) and Fig. S3].

We identified the ionization phase transition of ammonia using high-pressure Raman spectroscopy up to 200 GPa [Fig. 2(a)]. The significant changes in the low-frequency (< 1100 cm$^{-1}$) lattice modes [Fig. 2(c)] and the emergence of new vibrational modes ($v_a$, $v_b$) in the high-frequency region (3000 - 3700 cm$^{-1}$) [Fig. 2(d)] are considered signatures of the phase transition from phase V to phase β. The observed changes agree with previous reports. The reported phase transition pressure of Ninet *et al*. and Palasyuk *et al*. are ~150 GPa and ~130 GPa respectively [16,17]. Similarly, we used the appearance of $v_b$ mode as evidence of the phase transition, with the transition pressure in the experiment shown in Fig. 2 being ~ 150 GPa. In other runs, the appearance of $v_b$ mode was also observed at ~ 135 GPa [Fig. S2]. The pressure gradient in the sample chamber or systematic error from the diamond edge could be the reason for the variation in the phase transition points.

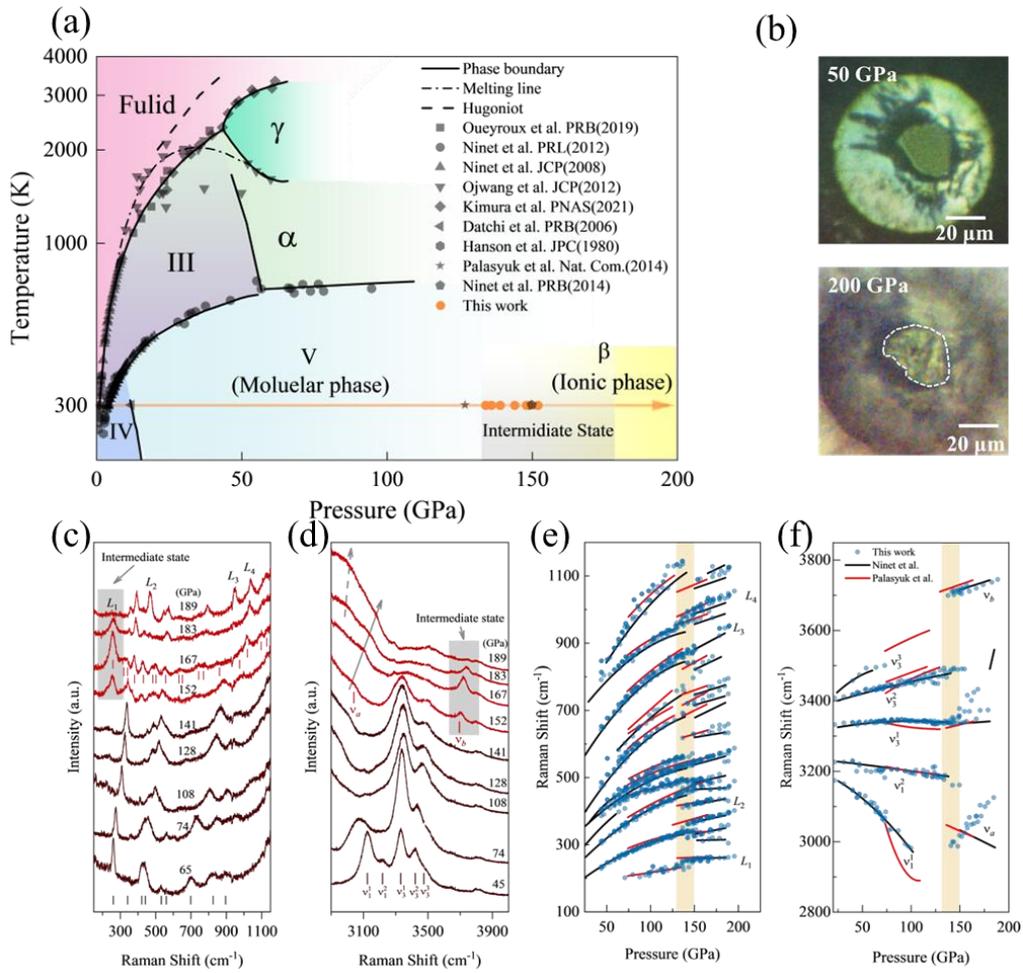

Fig. 2. (a) Phase diagram of NH$_3$ and the experimental path (orange arrow) in this work. The solid black lines are the phase boundaries determined from Ref. [16,17,24,25,28,31–34]. The dashed line represents the Hugoniot curve [35]. The dashed-dotted lines represent the melting curve [31,33]. The colored solid circles represent the ionization phase transition pressures in seven independent runs. (b) Optical image of the sample chamber. The dashed line is a guide for the eye to the sample chamber. (c) and (d) The representative Raman spectra of ammonia at different pressures. The red and brown solid lines represent Raman spectra of phases β and V, respectively. The solid bar represents the identified Raman modes. The gray solid arrow denotes the identified $v_a$ mode in this work. The dashed gray arrow indicates $v_a$ mode reported in previous studies. Pressure dependence of lattice modes (e) and vibrational modes (f). The orange area represents the V-β phase transition range. The red and black solid lines are the data from Ref. [17] and Ref. [16]. The blue solid circles represent the data measured in our experiments, including a total of seven runs.

All Raman modes have been identified. Overall, the lattice modes exhibit consistent pressure-dependent behavior with previous work [Fig. 2(e)]. The $L_1$ and $v_b$ modes serve as the most direct evidence of the phase transition. After the phase transition to phase β, $L_1$ and $v_b$ modes exhibit anomalous behavior [Fig. 2(d)]. Notably, these two modes are absent from the theoretically predicted stable structure of the $Pma2$ space group for ionic ammonia. We observed the simultaneous disappearance of $L_1$ and $v_b$ modes [Fig. 2(c) and 2(d)]. The trend of disappearance of the $v_b$ and $L_1$ modes was also observed in another run (Fig. S3). This confirms that pressure induces the disappearance of the $L_1$ and $v_b$ modes. We analyzed the behavior of the two Raman modes prior to their disappearance in detail (Fig. S1). The full width at half maximum (FWHM) of $L_1$ and $v_b$ did not show significant changes before the disappearance [Fig. S1(c)]. In contrast to $L_1$ mode, the intensity of $L_4$ lattice mode did not exhibit a weakening trend [Fig. S2(b)]. At the same time, the intensity of $L_2$ and $L_3$ modes were found to be enhanced [Fig. 2(c)]. The relative intensity of $L_1$ and $v_b$ shows almost the same trend with pressure, indicating a strong correlation between the $L_1$ and $v_b$ modes [Fig. S2(b)].

The $v_a$ mode was also identified as the characteristic peak of phase β. We noticed that the pressure dependence of the $v_a$ mode is inconsistent with previous reports. This can be attributed to the relatively weak signal of $v_a$ and its interference from the second-order peak of diamond. The $v_a$ peak identified from our fitting is indicated by solid gray arrow [Fig. 2(d)]. We noticed the presence of a shoulder on the left side. Visually, this shoulder exhibits a redshift trend consistent with previous reports (dashed gray arrow). It could be the $v_a$ identified in previous works. However, it could not be accurately fitted. The distinction between blue and redshifts of $v_a$ does not affect the identification of the phase β. More importantly, $v_a$ and $v_b$ exhibit completely different behaviors. No significant change was observed in the Raman spectra near the $v_a$ region after the disappearance of $v_b$. We observed the same phenomenon in a laser heating experiment [Fig. S4]. We stopped increasing pressure when $L_1$ and $v_b$ were clearly observed. $L_1$ and $v_b$ also disappear after the low-power, short-time laser heating. However, the Raman spectrum near the $v_a$ region remained unchanged. For solid molecules, variations in vibrational modes reflect modifications of molecular configurations. It is not easy to disrupt the molecular configuration. Laser heating above 2000 K still has no significant effect on the vibrational mode of phase V [31,33]. Considering that both pressure and temperature can induce the disappearance of $v_b$, it is likely that $v_b$ does not originate from the vibrational modes of ionic ammonia as previously thought.

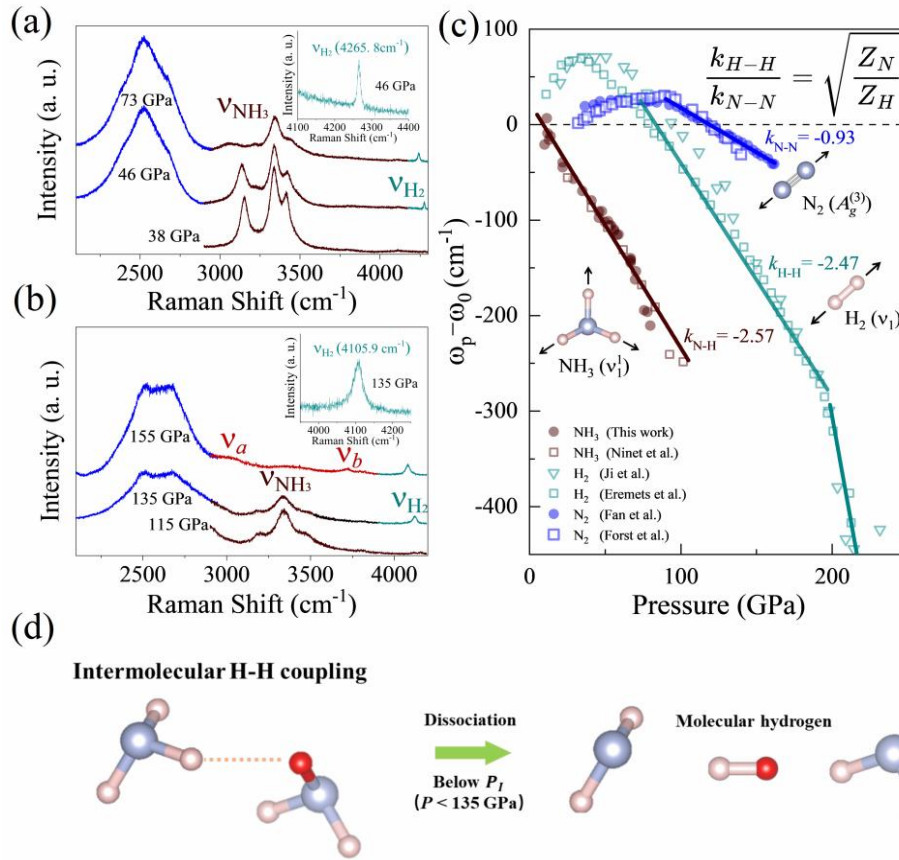

Fig. 3. (a) and (b) The Raman spectra of observed hydrogen vibrational modes in two independent runs. The blue broad peak is the second-order Raman peak of diamond. The brown line represents the vibrational modes of molecular ammonia (phase V). The red line represents the vibrational modes of ionic ammonia (phase β). The green line represents the vibrational mode of hydrogen. (c) The symmetry stretching modes as a function of pressure in dense hydrogen ($v_1$), nitrogen ($A_g^{(3)}$), and ammonia ($v_1^1$). The $\omega_0$ and $\omega_p$ represent the frequency of the vibrational mode at zero pressure, and corresponds to the frequency at the given pressure. The solid symbols represent our data. The hollow symbols represent data from the Ref. [16,30,36,37]. The $k$ represents the slope of the Raman frequency versus pressure curve, and $Z$ represents the atomic number. (d) The schematic

diagram illustrating the formation of molecular hydrogen based on intermolecular H-H coupling below ~ 135 GPa (ionization phase transition pressure, $P_I$).

The molecular-to-ionic phase transition was proposed to proceed through two stages: (i) interplane proton transfer, and (ii) structural rearrangement [17]. Considering the limited ability of S-XRD to characterize hydrogen atoms [29,30], we infer that the disappearance of $L_1$ and $v_b$ modes has a close relationship with the behavior of hydrogen atoms. Other phenomena observed suggest the involvement of hydrogen. In a few experiments, we observed sporadic molecular hydrogen [Fig. 3(a) and 3(b)]. The hydrogen Raman spectra were observed at ~ 40 GPa and ~ 150 GPa in two runs. The formation of hydrogen in ammonia after laser heating above 2000 K is a common phenomenon (Eq. 1).

$$2NH_3 \xrightarrow{>2000\ K} N_2 + 3H_2 \quad (1)$$

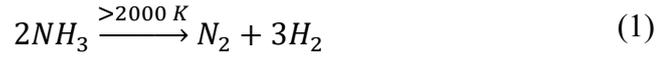

However, no heating methods were employed in this study. Ninet *et al*. reported observation of hydrogen signals at 150 GPa without signals of nitrogen due to the decomposition of ammonia under X-ray irradiation [14]. The absence of nitrogen signals is attributed to nitrogen being in an amorphous phase at this pressure [42]. In this study, if the appearance of hydrogen is attributed to the decomposition of ammonia, the signal of nitrogen would also be expected to be observed. The absence of detectable nitrogen signals may be attributed to their weak intensity, which is overwhelmed by the second-order Raman peak of diamond. We rule out this possibility. The hydrogen observed by Oueyroux *et al*. using laser heating above 2000 K, at 30 GPa was much weaker than that detected in this study [31]. However, weak nitrogen signals were observed near the second-order Raman mode of diamond. Therefore, hydrogen molecules must have been generated from ammonia through a different mechanism. In the absence of molecular nitrogen formation, the only possible pathway is dissociation in to hydrogen molecules as follows:

$$2NH_3 \xrightarrow{300\ K} 2NH_2 + H_2 \quad (2)$$

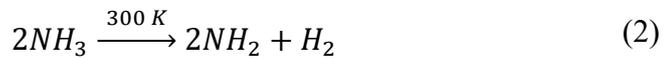

Eq. 2 implies the possible existence of intermolecular H-H coupling in ammonia.

A comparative analysis of the soft vibrational modes was performed to elucidate intermolecular interactions in ammonia. The softening of vibrational modes with increasing pressure is a common phenomenon in molecular crystals, reflecting enhanced intermolecular interactions. Similar to phase V ammonia, the $\lambda$-$N_2$ and I-$H_2$ phases remain stable over wide pressure ranges of approximately 30–170 GPa and 0–200 GPa, respectively [36,38,39]. Hydrogen and nitrogen molecules consist of a single element, and their vibrational modes are influenced solely by the intermolecular interactions among identical atoms. In the pressure range where the anharmonic effect weakens after the vibrational mode's turning point, a linear function was used to fit the pressure dependence of vibrational frequency for $H_2$-$\nu_1$, $N_2$-$A_g^{(3)}$ and $NH_3$-$\nu_1^1$ [Fig. 3(c)]. The fitted slopes are $k_{N-H} = -2.57$, $k_{H-H} = -2.47$ and $k_{N-N} = 0.93$. Notably, the value of $k_{N-N}/k_{H-H}$ is 0.377. The value of $\sqrt{Z_H/Z_N}$ is 0.376 ($Z_H = 1$, $Z_N = 7$). Their relationship can be well described by an equation, $\sqrt{Z_H/Z_N} = k_{N-N}/k_{H-H}$. The redshift trend of the vibrational mode is related to the atomic number. Coulomb interactions, one of the fundamental forms of interparticle interaction, provide a qualitative explanation for the dependence on atomic number as:

$$\vec{F} = \frac{1}{4\pi\varepsilon_0} \frac{q_1 q_2}{r^2} \vec{e}_r \tag{3}$$

Where $\varepsilon_0$ represents the vacuum permittivity, $q_1$ and $q_2$ represent the electric charges, $r$ represents the distance between two particles. Considering that the atomic charge ($q$) is related to $Z_e$, the strength of the Coulomb interaction is also related to $Z_e$. Noting that $k_{H-H} \approx k_{N-H}$, the interactions in hydrogen molecules occur solely between hydrogen atoms. It indicated the presence of H-H interactions between ammonia molecules. Therefore, the molecular hydrogen originates from dissociation through H–H coupling between ammonia molecules [Fig. 3(d)].

Theoretical calculations were performed to give an insight into intermolecular interaction in dense ammonia. The calculated equation of state of ammonia agrees well with experimental data (Fig. S6). The electron localization function (ELF) of phase IV ammonia was calculated based on static structure. The isosurfaces of 0.5-value was defined as that of a free electron [40]. At 60 GPa, ELF isosurfaces start to connect between N and H atoms (1.67 Å apart), indicating the onset of intermolecular coupling. This pressure is close to the lowest pressure that of sporadic hydrogen observed [Fig. 2(a)]. Under static structure, it cannot reflect the actual behavior of hydrogen. We performed AIMD to investigate the temporal evolution of hydrogen behavior. With increasing pressure, maximum mean square displacements (MSDs) of H atoms rise obviously after 60 GPa [Fig. 4(c)]. This pressure is close to that where a slight change in Raman spectra, which identified as Phase VI [27]. The MSDs of nitrogen showed no significant change. This explains why the Phase VI was not supported by the S-XRD. The atomic trajectories show an increase in hydrogen disorder, especially at 120 GPa [Fig.4(b)]. The increasing disorder is also consistent with the behavior of a vibrational mode ($v_3^1$). The FWHM of $v_3^1$ exhibits a liner increasing trend prior to the phase transition, with a turning point about 70 GPa [Fig. S7(c)]. We calculated the time dependence of H-H distances between two ammonia molecules. The shortest H-H distances below 1.67 Å (the distance at which ammonia molecules start to couple) at all calculated pressures [Fig. 4(c)], indicating the possible occurrence of intermolecular H–H coupling. Because H–H coupling occurs only under specific molecular orientations, the hydrogen signal is not always observed below ~ 135 GPa.

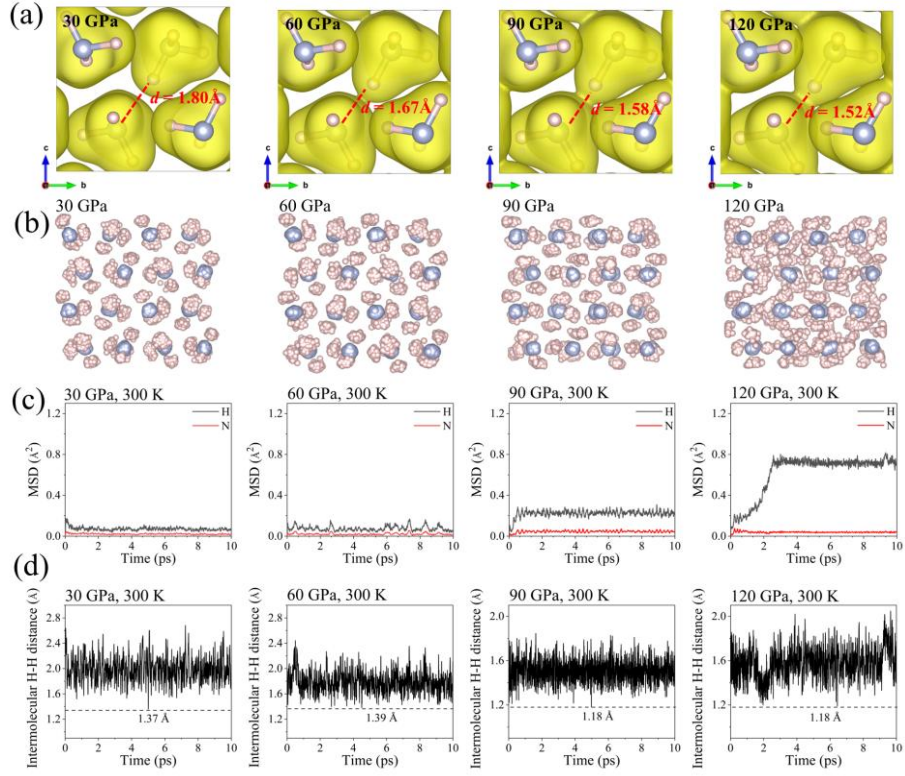

Fig. 4. (a) Electron localization function (ELF) isosurfaces of 0.5 plotted at different pressures, calculated using the static structure. Red dashed lines indicate the distance between two NH$_3$ molecules. The NH$_3$ molecules' ELF isosurfaces start to connect at 60 GPa. (b) The atomic trajectories of the atomic positions at different pressures. (c) Calculated mean square displacements (MSD) from *ab initio* molecular dynamics (AIMD) at different pressures. (d) The time dependence of intermolecular distance between the nearest two NH$_3$ molecules at different pressures. Black dashed lines are guides for eye to the minimum distance between two intermolecular hydrogen atoms.

The H-H coupling gives rise to two different experimental phenomena [Fig. S9]. At pressure below ~ 135 GPa, it leads to dissociation, giving rise to molecular hydrogen. At pressure above ~ 135 GPa, it serves as an intermediate state in the ionization of ammonia. Hence, $v_a$ and $v_b$ correspond to two pathways. Both can lead to the ionization of ammonia. $v_b$ associated with H-H coupling (green arrow) (FIG. 1). The frequency of $v_b$ satisfy the inequation $v_1^1(NH_3) < v_b < v_1(H_2)$. At ~ 180 GPa, the frequency of $v_1$ mode is about 3850 cm$^{-1}$, while the $v_b$ mode is about 3750 cm$^{-1}$. In the proposed *Pma*2 structure, the bond length of the N-H is approximately 1.01 Å. The H-H bond length in Phase IV hydrogen is approximately 0.83 Å [38]. Unreasonably, the $v_b$ mode originating from N-H vibrations is so close to the mode of H-H vibrations. The $v_b$ mode is more likely to originate from a H-H coupling. The disappearance of the $v_b$ mode can be attributed to the effect of pressure or temperature on this intermediate state. This intermediate state will eventually transform into ionic ammonia. Thus, $v_a$ is not change. More importantly, it explains why the $L_1$ and $v_b$ modes are absent in the Raman spectrum calculated from the stable $Pma$2 structure. Notably, the $L_1$ is also very similar to that of dense hydrogen in phase IV. When the dense hydrogen enters phase IV, a lattice mode appears around 300 cm$^{-1}$. It originates from coupling with $H_2$ molecules [38,41]. We must acknowledge that, due to the instability of this intermediate state and strong quantum effect of hydrogen, an accurate spectrum calculation is very challenging [42,43].

In a conclusion, we systematically investigated the Raman spectra of dense ammonia up to 200 GPa, combined with theoretical calculations. Our results reveal that hydrogen bonding is not the only intermolecular interaction present in dense ammonia at 300 K. We reveal a novel dissociation pathway of ammonia and a dual-path mechanism in ammonia's ionization. However, directly characterizing the behavior of hydrogen remains a significant challenge under high pressure. The behavior of hydrogen atoms remains an open question in hydrogen-containing systems. The presence of $H_2$ molecules resulting from intermolecular H–H coupling suggests that the potential existence of a small amount of hydrogen should be considered in planetary ices composed of $CH_4$, $H_2O$, and $NH_3$.

**Acknowledgments**

This work was financially supported by the National Natural Science Foundation of China (Grant No. 12374013) and the Fundamental Research Funds for the Central University (Grant No. 2020SCUNL107).

# Supplementary Material for
# "Local Proton Disorder Induced Intermolecular H-H Coupling in Ionization of Dense Ammonia"

Yu Tao, Li Lei*, Jingyi Liu and Binbin Wu

*Institute of Atomic and Molecular Physics, Sichuan University, 610065 Chengdu, China*

*Electronic mail: lei@scu.edu.cn

**In-situ high-pressure Raman Scattering**

Ultra-high pressures were generated using a diamond anvil cell (DAC) with a minimum culets size of 80 μm. Other culet sizes, such as 100 μm, were also employed. A rhenium gasket was employed, which was pre-compressed to a thickness of 4 μm. A sample chamber with approximately 25 μm diameter was drilled using laser cutting techniques. Then, the assembled DAC was placed in a nitrogen-filled glovebox, and cooled to approximately 100 K with liquid nitrogen. Gaseous ammonia was introduced into the glovebox through a sealed gas pipeline and applied to the surface of the sample chamber. Once the gaseous ammonia contacted cooled DAC, it rapidly liquefied. After the sample chamber was filled with liquid ammonia, the DAC was sealed and the pressure was increased to 10 GPa. Raman spectra were collected using a custom-built confocal Raman spectrometry system in a back-scattering geometry, equipped with an Andor Shamrock triple-grating monochromator and an Andor Newton Electron Multiplying Charge Coupled Device (EMCCD). A 532 nm solid-state laser was used as the excitation source in this study. The pressure was calibrated using the edge of the first-order Raman peak of the diamond [1–3]. A total of seven runs were carried out in this study.

**Laser heating experiment (FIG. S3)**

In this run, we stopped increasing the pressure when the signals of $L_1$ and $v_b$ were clear detected. The sample was re-examined on the next day. The intensities of the $L_1$ and $v_b$ signals showed no reduction (the Raman spectrum at 154 GPa). Due to diamond relaxation, the chamber pressure increased by ~3 GPa. It confirms that the observed disappearance of the $L_1$ and $v_b$ modes was caused by the increasing pressure. Then, the sample chamber was heated using a 1064nm laser with a power of approximately 20 W. The heating time was within 30s. At about 150 GPa, ammonia remains optically transparent. The heating efficiency of the 1064 nm is relatively low. Hence, the temperature is not high. After heating, the $L_1$ and $v_b$ disappeared, while a boarding $v_a$ vibrational mode signal was still detectable. In previous work, Queyroux et al. heated molecular ammonia up to 3000 K at 31 GPa by $CO_2$ laser [4]. After heating, no change was observed in the vibrational mode signal of solid molecular ammonia. Phase V ammonia can remain stable under high-temperature and high-pressure conditions. The disappearance of the vibrational mode signal requires the breakdown of the molecular structure, which is, however, not easily achieved. However, the $v_b$ mode still disappeared after heating. It suggests that the $v_a$ and $v_b$ cannot both be attributed to Raman modes of ionic ammonia.

**Computational methods**

First-principles density functional theory (DFT) calculations and *ab initio* molecular dynamics (AIMD) simulations were performed using the Quantum ESPRESSO (QE) codes [5]. Atomic nuclei were treated as classical nuclei. The quantum nature of the proton is not taken into account. We use the Perdew-Burke-Ernzerhof (PBE)-type generalized gradient approximation (GGA) for the exchange-correlation functional and the projector-augmented wave (PAW) (energy cutoff of 80 Ry). Convergence tests with a threshold of 1meV per atoms in energy to a Monkhorst-Pack *k*-point grid of 9×5×6. We performed variable cell relaxations (lattice parameters and atomic positions) on the experimental structures of Phase V ammonia at 24 GPa confirmed by neutron diffraction reported by S. Ninet [6]. The optimized structures at different pressure were fitted well with high-pressure single-crystal x-ray diffraction experiments data reported by F. Datchi [7]. The electron localization function (ELF) was plotted using VESTA software [8]. Phonon calculation and AIMD simulations were both performed on a 2×2×2 supercell method (32 $NH_3$ molecular). Phonon calculation was performed using the Phonopy package, in which the finite-displacements method is implemented [8]. AIMD simulations were performed in the canonical (NVT) ensemble. Each simulation consists of 10 000 time steps with a time step of 1 fs.

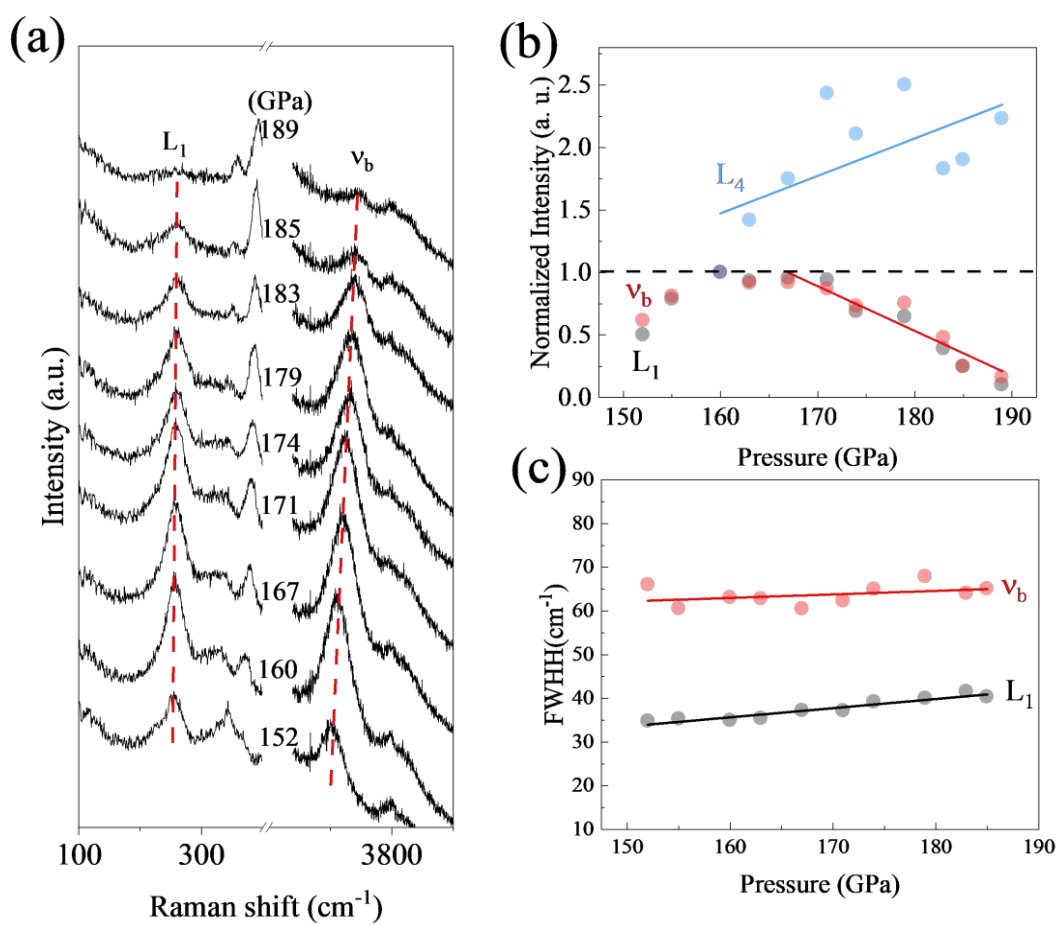

Fig. S1. (a) Raman spectra of $L_1$ and $\nu_b$ vibrational modes as a function of pressure. (b) The normalized intensity of $L_1$, $\nu_b$ and $L_4$ modes as function of pressure. (c) The full width at half maximum (FWHM) of $L_1$ and $\nu_b$ modes as a function of pressure.

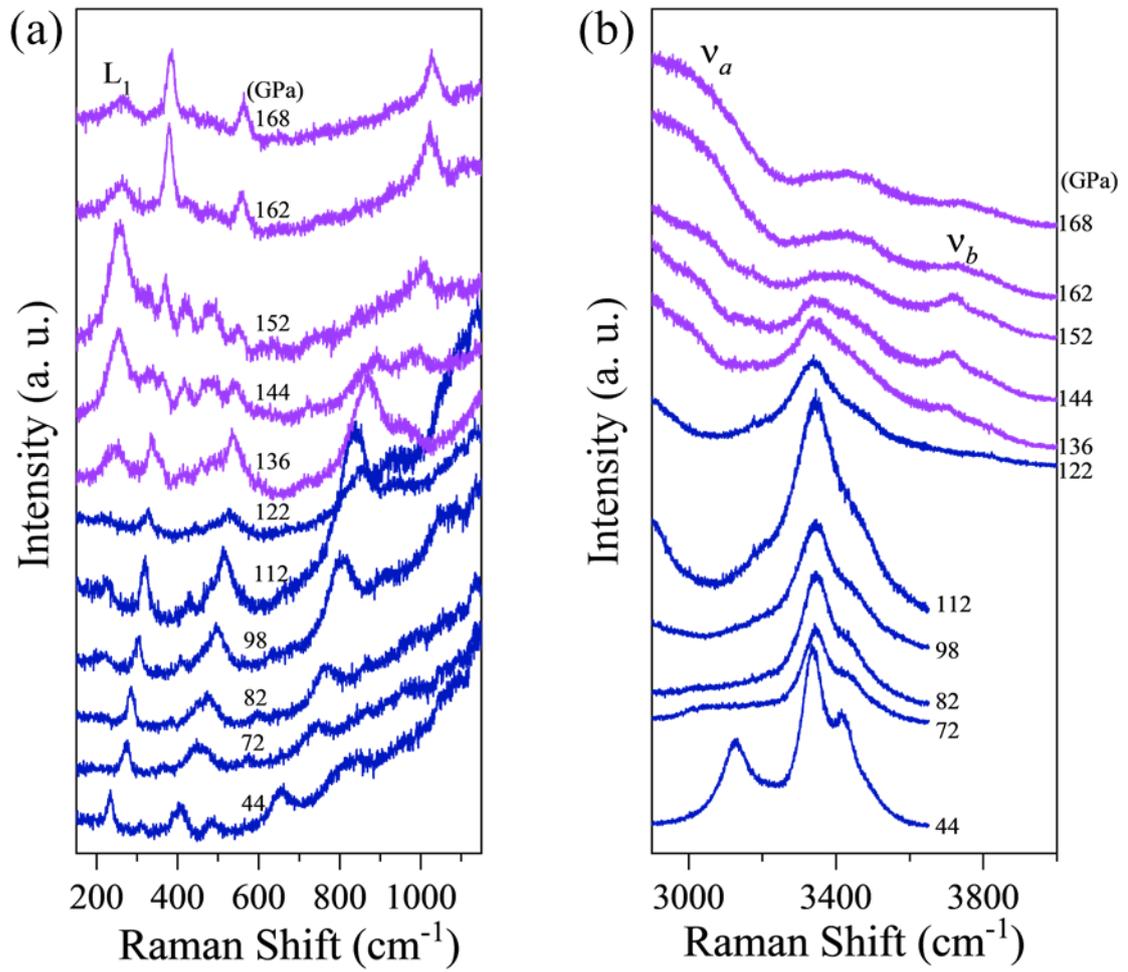

Fig. S2. The L$_1$ (a) and ν$_b$ (b) modes disappear with increasing pressure after phase transition observed in other experiments we conducted.

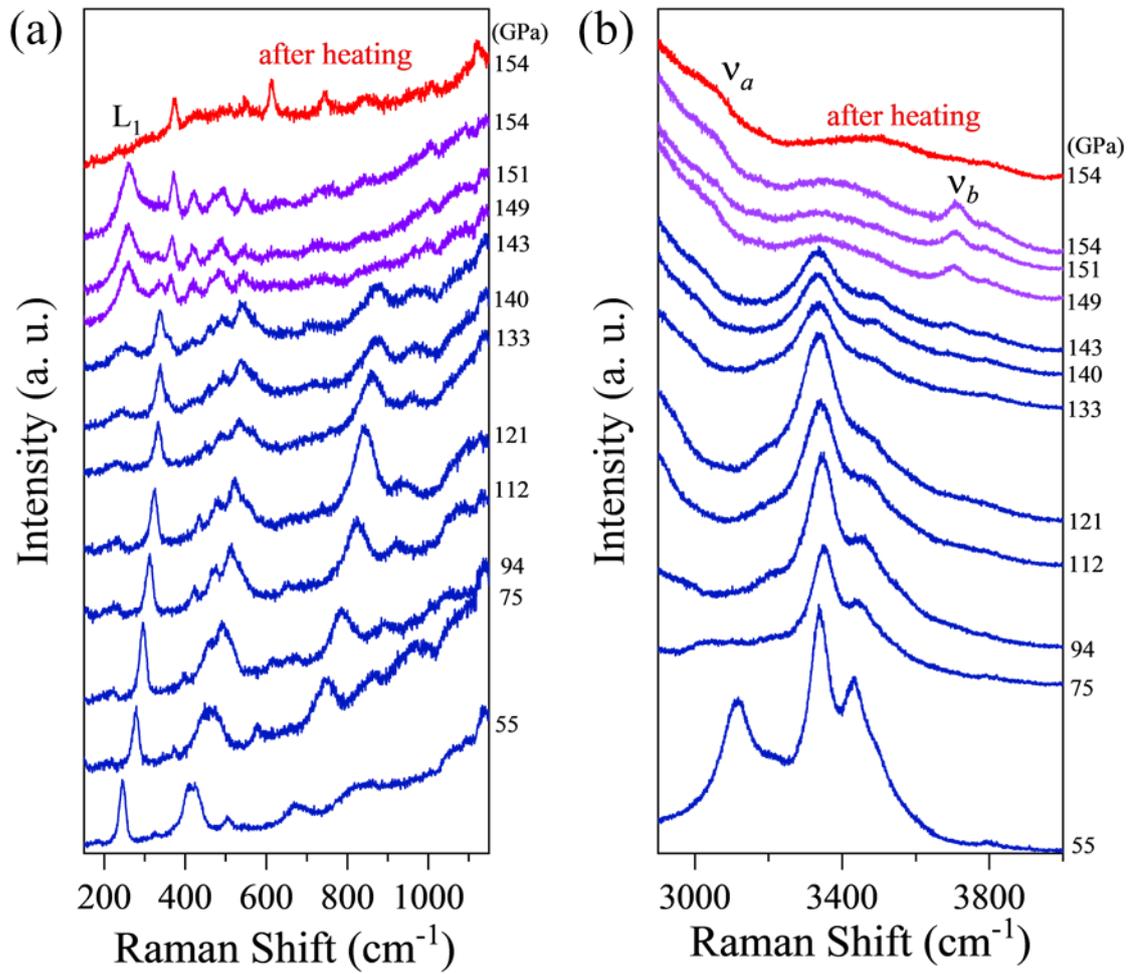

Fig. S3. The disappearance of the $L_1$ (a) and $\nu_b$ (b) Raman modes induced by 1064 nm laser heating. The red lines are the Raman spectra after lasering heating.

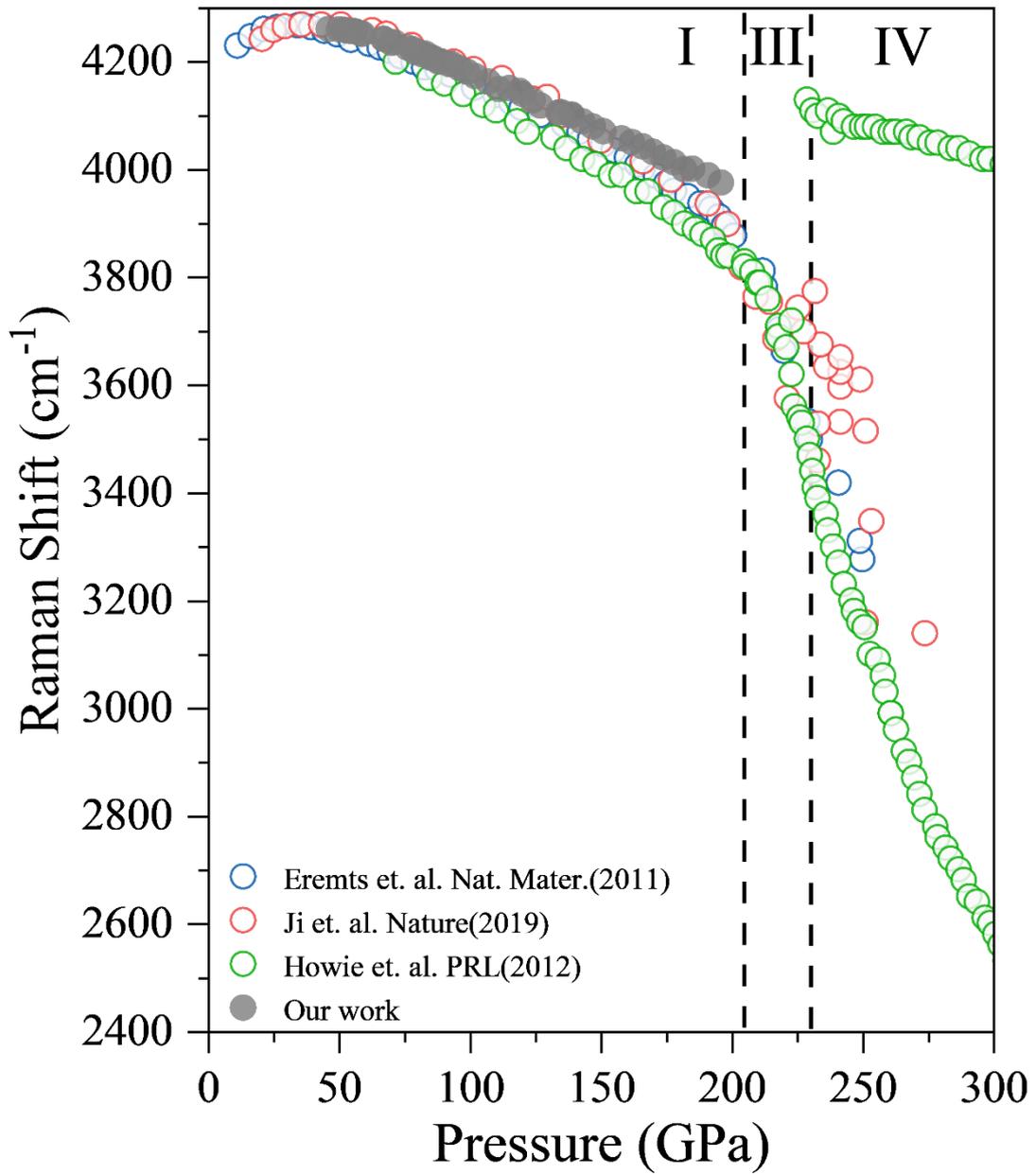

Fig. S4. The pressure dependence of hydrogen vibrational mode dissociated from ammonia in this study, compared to that of pure hydrogen [10–12].

Table S1. The structure information of Phase V NH$_3$. The structure determined by neutron diffraction at 24 GPa was used [6].

| | | NH$_3$ (Phase V) | |
|---|---|---|---|
| | | **Exp.** | **Calc. [This work]** |
| Pressure (GPa) | | 24 | 30 |
| Space group | | *P*2$_1$2$_1$2$_1$ | *P*2$_1$2$_1$2$_1$ |
| Z | | 16 | 16 |
| a (Å) | | 2.9215(5) | 2.85323(9) |
| b (Å) | | 5.0921(11) | 4.97439(3) |
| c (Å) | | 4.8056(5) | 4.73655(9) |
| α=β=γ (°) | | 90 | 90 |
| V (Å$^3$) | | 71.4908(49) | 67.2269(6) |
| **Atomic positions** | | Fractional atomic coordinates (x, y, z) | |
| Atom | Wyckoff Position | **Exp.** | **Calc.** |
| N | 4a | 0.2388(22), 0.3363(15), 0.2552(13) | 0.24005(3), 0.34678(8), 0.25486(2) |
| H | 4a | 0.3508(22), 0.1415(23), 0.2321(14) | 0.36534(5), 0.15715(0), 0.22234(6) |
| H | 4a | -0.0823(40), 0.3145(21), 0.3226(21) | 0.90454(4), 0.32752(1), 0.32559(5) |
| H | 4a | 0.2550(61), 0.4559(22), 0.0731(20) | 0.28204(8), 0.55940(8), 0.56159(8) |

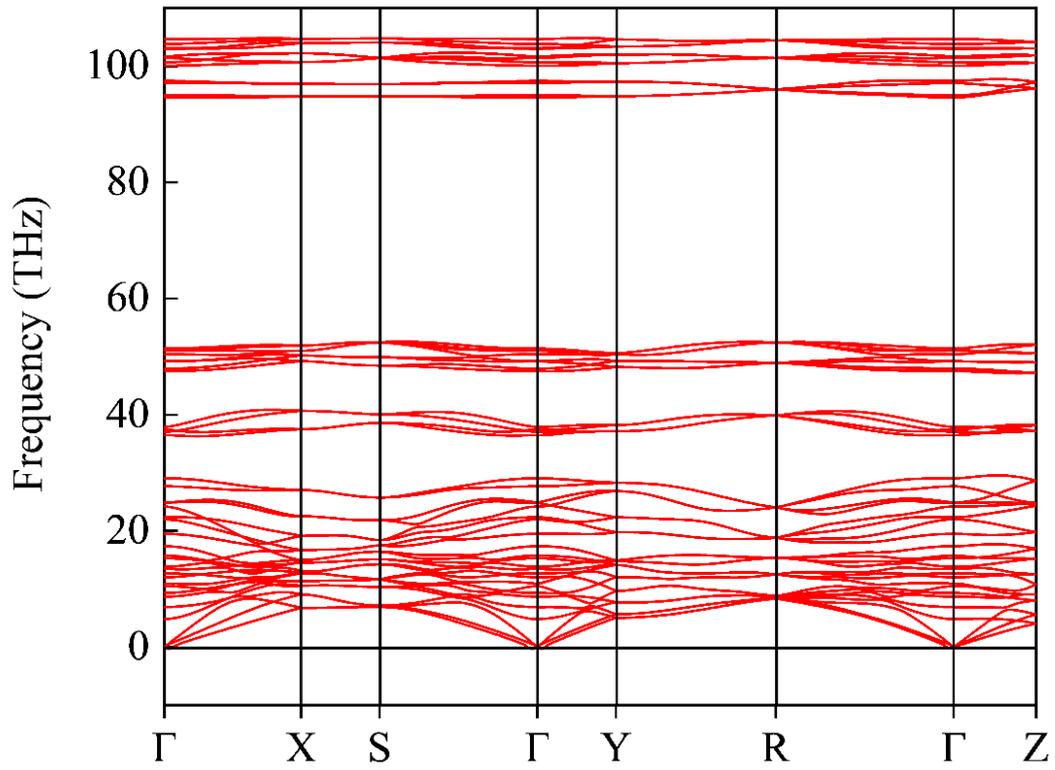

Fig. S5. Calculated phonon dispersion of Phase IV $NH_3$ at 30GPa.

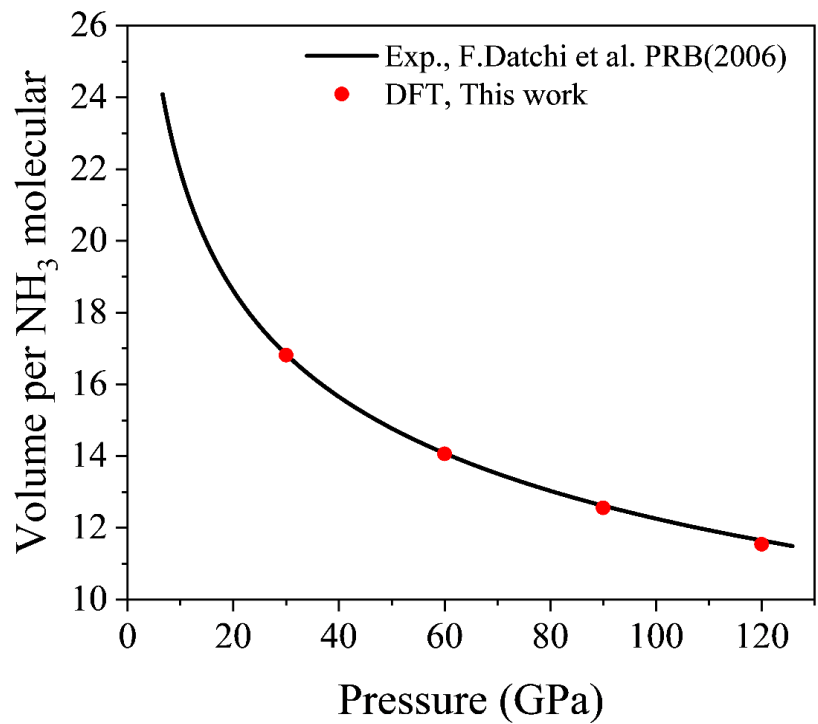

Fig. S6. Comparison of the volume per NH$_3$ molecular from DFT variable-cell calculations with experimental data [7]. The experimental data was fitted by Vinet equation.

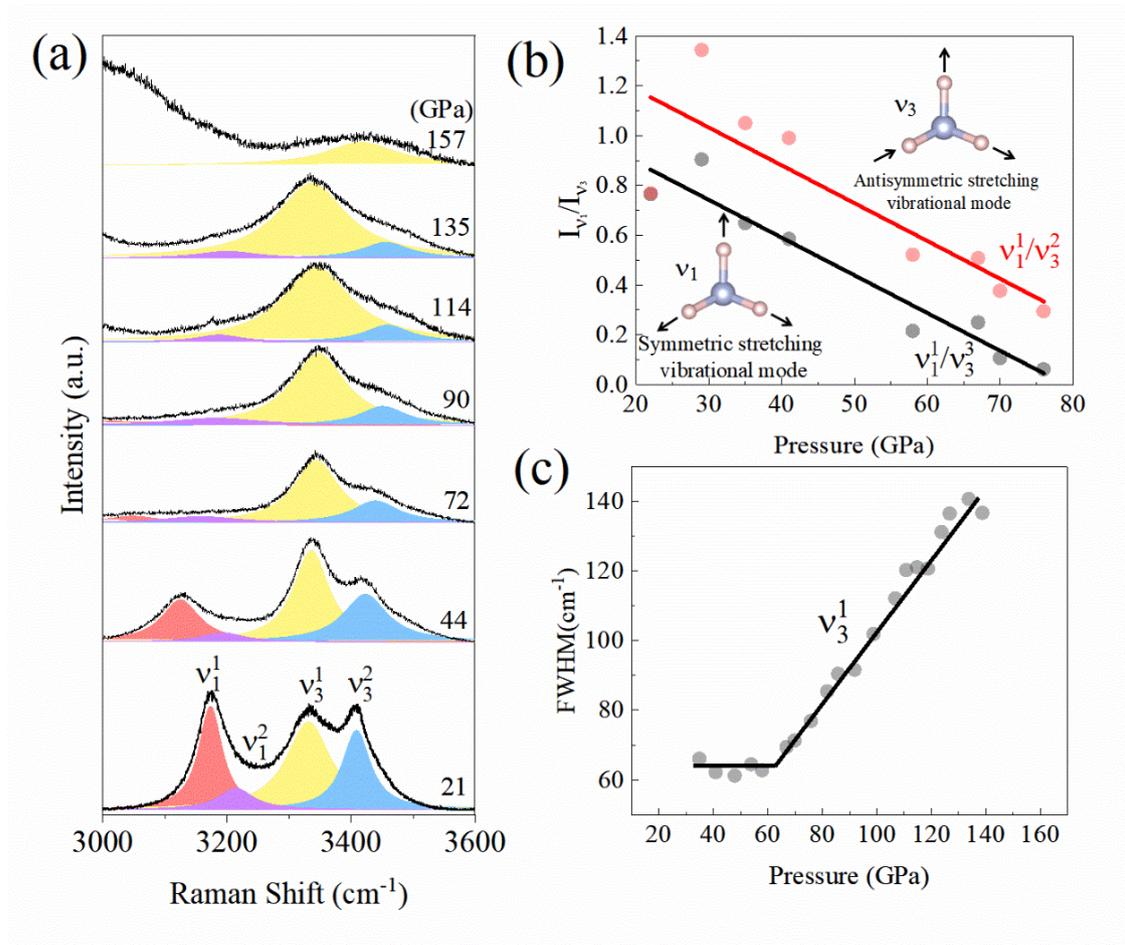

Fig. S7. (a) Lorentzian fitting of vibrational modes at different pressures. (b) The intensity ration of $v_1$ and $v_3$ as a function of pressure. The inset is schematic diagrams of the symmetric ($v_1$) and asymmetric ($v_3$) stretching vibrational modes of ammonia molecules. (c) The full width at half maximum (FWHM) of $v_3^1$ mode as function of pressure.

Fig. S9. Two experimental phenomena could be induced by intermolecular H-H coupling. At pressure above ionization phase transition pressure ($P_I$) ($P > 135$ GPa), it drives the ionization of ammonia. At pressure below $P_I$ ($P < 130$ GPa), it leads to dissociation, giving rise to sporadic molecular hydrogen.